# Phase estimation by photon counting measurements in the output of a linear Mach Zehnder (MZI) interferometer

**Ben-Aryeh**

Department of physics, Technion-Israel Institute of Technology, Haifa, 32000 Israel

Corresponding author: phr65yb@ph.technion.ac.il

Photon counting measurements are analyzed for obtaining a classical phase parameter in linear Mach Zehnder interferometer (MZI), by the use of phase estimation theories. The detailed analysis is made for four cases: a) Coherent states inserted into the interferometer. b) Fock number state inserted in one input port of the interferometer and the vacuum into the other input port. c) Coherent state inserted into one input port of the interferometer and squeezed-vacuum state in the other input port. d) Exchanging the first beam-splitter (BS1) of a MZI by a non-linear system which inserts a NOON state into the interferometer and by using photon counting for parity measurements. The properties of photon counting for obtaining minimal phase uncertainties for the above special cases and for the general case are discussed.
*OCIS codes: 350.5030, 040.5570, 350.5730,*

## 1. INTRODUCTION

Following general measurement theories [1-4] phase can be considered as a classical parameter so that phase measurements and their uncertainties can be related to 'quantum estimation theories'. In the present article we apply this approach for treating a linear Mach Zehnder (MZI) interferometer in which two-mode electromagnetic (EM) fields are entering into the interferometer through an 'ordinary' first beam-splitter (BS1), a phase difference $\varphi$ is inserted between the two arms of the interferometer, and the two-mode EM fields are exiting through an 'ordinary' second beam-splitter (BS2). The role of MZI (referring also as the interferometer) is to measure the phase $\varphi$ and we study the application of the phase estimation methods for this purpose.

In optical interferometry, the resource for phase estimation is identified to be the number of photons $N$ to reach a desired precision. Classically, the precision of the estimated phase scales as $1/\sqrt{N}$, the so called standard quantum limit (SQL). By employing *entangled states* the precision can be improved to scale as $1/N$ known as the Heisenberg limit [5-13]. We would like to study the limits obtained for phase estimation for various measurements on the output fields of MZI.

The operators which form an SU(2) rotation group, and describe MZI are given as

$$\hat{J}_x = (\hat{a}^\dagger \hat{b} + \hat{b}^\dagger \hat{a})/2 \quad ; \quad \hat{J}_y = (\hat{a}^\dagger \hat{b} - \hat{b}^\dagger \hat{a})/2i \quad ; \quad \hat{J}_z = (\hat{a}^\dagger \hat{a} - \hat{b}^\dagger \hat{b})/2, \quad (1)$$

where $\hat{a}$ and $\hat{b}$ are the two mode operators which satisfy bosonic commutation relations (CR), $[\hat{a}, \hat{a}^\dagger] = [\hat{b}, \hat{b}^\dagger] = 1$. The total photon number operator is $\hat{N} = \hat{a}^\dagger \hat{a} + \hat{b}^\dagger \hat{b}$ and

$$\hat{J}^2 = \hat{J}_x^2 + \hat{J}_y^2 + \hat{J}_z^2 = (\hat{N}/2)(\hat{N}/2 + 1) \quad , \quad (2)$$

where the eigenvalues of $\hat{J}^2$ are j(j+1).

Two-mode state $|\psi\rangle_{in}$ inserted into the MZI is transformed by the first beam-splitter (BS1) as

$$|\psi\rangle = \hat{U}_{BS1} |\psi\rangle_{in} \quad , \quad (3)$$

where $\hat{U}_{BS1}$ represents the unitary transformation made by BS1. The general EM state obtained after BS1 can be described as

$$|\psi\rangle = \sum_{j=0}^{\infty} \sum_{m=-j}^{j} C_{j,m} |j,m\rangle \quad , \quad (4)$$

where $|j,m\rangle$ is an eigenstate of $\hat{J}^2$ and $\hat{J}_z$. The explicit expression for the amplitudes $C_{j,m}$ depends on the choice of $|\psi\rangle_{in}$ and $\hat{U}_{BS1}$ so that in the meantime the analysis is general. Quite often it is convenient to represent the state $|\psi\rangle$ as a superposition of number states

$$|\psi\rangle = \sum_{n(1),n(2)=0}^{\infty} C_{n(1),n(2)} |n(1),n(2)\rangle \quad (5)$$

of the two EM modes. Such representation can be transformed into the form of Eq. (4) (or vice versa Eq. (4) can be transformed into the form of Eq. (5) ) by using the relations

$$\frac{n(1) + n(2)}{2} = j \quad ; \quad \frac{n(1) - n(2)}{2} = m \quad . \quad (6)$$

The representation (4) is convenient for phase estimation in MZI as a phase difference inserted between the two arms of the interferometer is transforming the state $|\psi\rangle$ into

$$|\tilde{\psi}\rangle = \exp(-i\hat{J}_z \varphi)|\psi\rangle = \sum_{j=0}^{\infty} \sum_{m=-j}^{j} C_{j,m} \exp(-im\varphi) |j,m\rangle \quad . \quad (7)$$

The effect of the second beam-splitter (BS2) in MZI on $|\tilde{\psi}\rangle$ is given as

$$|\psi\rangle_{out} = \hat{U}_{BS2}|\tilde{\psi}\rangle \quad , \tag{8}$$

where $\hat{U}_{BS2}$ represents the unitary transformation obtained by BS2.

The usual approach in phase estimation is to fix a particular phase dependent observable $\hat{O}$ at the output and look for its behavior under the phase change $\varphi$. For the present general case of MZI we get

$$\langle \hat{O} \rangle = \langle \psi|_{out} \hat{O} |\psi\rangle_{out} = \langle \tilde{\psi}|\hat{U}^{\dagger}_{BS2}\hat{O}\hat{U}_{BS2}|\tilde{\psi}\rangle = \langle \tilde{\psi}|\hat{O}_{in}|\tilde{\psi}\rangle \quad , \tag{9}$$

where we have defined

$$\hat{O}_{in} = \hat{U}^{\dagger}_{BS2}\hat{O}\hat{U}_{BS2} \quad . \tag{10}$$

We find that while the observable $\hat{O}$ is given at the output of the interferometer, the observable $\hat{O}_{in}$ is given inside the interferometer, before BS2.

For phase estimation the phase uncertainty is typically given as

$$\delta\varphi = \frac{\Delta\hat{O}}{|\partial/\partial\varphi\langle\hat{O}\rangle|} \quad ; \quad \Delta\hat{O} = \sqrt{\langle\hat{O}^2\rangle - \langle\hat{O}\rangle^2} \quad . \tag{11}$$

Using the above relations the expectation values needed in using Eq. (11) can be calculated by:

$$\langle\hat{O}\rangle = \langle\tilde{\psi}|\hat{O}_{in}|\tilde{\psi}\rangle \quad ; \quad \langle\hat{O}^2\rangle = \langle\tilde{\psi}|\hat{O}_{in}^2|\tilde{\psi}\rangle \quad . \tag{12}$$

By using Eq. (11) and a special form for $\hat{U}_{BS2}$ in Eq. (9) the calculations for expectation values can be simplified so that instead of calculating the expectation values $\langle\hat{O}\rangle$ and $\langle\hat{O}^2\rangle$ outside the interferometer one can calculate the equivalent expectation values $\langle\hat{O}_{in}\rangle$ and $\langle\hat{O}_{in}^2\rangle$ before BS2, i.e., with the wave function $|\tilde{\psi}\rangle$ of Eq. (7).

While there are many possibilities for choosing the observable $\hat{O}$, in the present work we analyze cases in which the observables are based on 'photon counting'. Photon counting measurements have been used by Noh, Fougeres and Mandel for phase estimation [14,15]. Their discussions were, however, on the use of operational methods for evaluating the *quantum phase*. The present approach is different, as it follows the phase estimation theories in which the wavefunction or more generally the density matrix is a function of the phase $\varphi$, considered as a classical parameter. Recently, photon counting methods known as 'Number resolving detectors' have been developed in relation to quantum information [16], interferometry [17-18], and parity measurements [19], among many others.

We do not consider here any nonlinear interactions produced inside the MZI, (e.g. [20-22]) but the source for the two-mode input states can be produced by any nonlinear interaction *before or instead of BS1*, (where "Instead of BS1" means that BS1 acts as a 'unit operator' after the non-linear interaction). We limit the discussion to phase estimation by photon counting on a linear interferometer.

A photon counting method can be described as follows: For quasi-steady EM beams the two outputs emerging from the interferometer can fall on two similar photodetectors which count the photoelectric pulses emitted during some time interval from $t$ to $t+T$. For EM pulses with a small number of photons we should carry the measurements over all the photons exiting the two outputs of the interferometer. The photon counting will exhibit fluctuations from one trial to the next. In each trial we can count the number of photons $l(1)$ and $l(2)$ exiting the first and second outports, respectively. In principle one can use photo-current differences in the interferometer output, but photon counting is more sensitive to fluctuations. Most important, in using photon counting we can calculate different photon statistics moments from the same experimental data.

We concentrate in the present article on the following two methods for phase estimation using photon photon counting :

**A. Phase estimation by output photon number difference ('First method')**

We calculate from photon counting the first and second moments given respectively by

$$\langle \hat{O} \rangle = \langle \hat{J}_z \rangle = \left\langle \left( \hat{a}^\dagger \hat{a} - \hat{b}^\dagger \hat{b} \right)_{out} \right\rangle \quad ; \quad \langle \hat{O}^2 \rangle = \langle \hat{J}_z^{\,2} \rangle = \left\langle \left( \hat{a}^\dagger \hat{a} - \hat{b}^\dagger \hat{b} \right)^2_{out} \right\rangle \quad , \quad (13)$$

where $\left( \hat{a}^\dagger \hat{a} \right)_{out}$ and $\left( \hat{b}^\dagger \hat{b} \right)_{out}$ are the number operators in the first and second output ports , respectively. By using many trials we can obtain $\langle \hat{O} \rangle = \langle \hat{J}_z \rangle$ representing the averaged difference in the number of photons, and $\langle \hat{O}^2 \rangle = \langle \hat{J}_z^{\,2} \rangle$ representing the averaged squared of photons numbers difference, in the two output ports. Experimentally, we can use these moments for phase estimation according to Eq. (11).

We simplify the calculations by assuming

$$\hat{U}_{BS2} = \exp\left(-i\frac{\pi}{2}\hat{J}_y\right) \quad . \tag{14}$$

Then by using Eq. (9) we get

$$\hat{O}_{in} = \hat{J}_x \quad ; \quad \hat{O}_{in}^{\,2} = \hat{J}_x^{\,2} \tag{15}$$

and according to Eq. (12) we get

$$\langle \hat{O}_{in} \rangle = \langle \tilde{\psi} | \hat{J}_x | \tilde{\psi} \rangle \quad ; \quad \langle \hat{O}_{in}^{\,2} \rangle = \langle \tilde{\psi} | \hat{J}_x^{\,2} | \tilde{\psi} \rangle \tag{16}$$

In the present approach we can choose the input state $|\psi_{in}\rangle$ and BS1 so that we get $|\psi\rangle$ of Eq. (4) using the transformation of Eq. (3) . By the insertion of a phase difference $\varphi$ between the two arms of the interferometer, $|\psi\rangle$ is transformed into $|\tilde{\psi}\rangle$ using Eq. (7). The photon counting measurements on the interferometer output state $|\psi\rangle_{out}$, given by Eq. (13), lead in a quite straightforward way to *equivalent* expectation values given by Eq. (16) over the state $|\tilde{\psi}\rangle$, where the phase uncertainty can be calculated by using Eq. (11).

It is quite straightforward to use this method for other states but one should notice that since operator $\hat{J}_x$ can lead only to $\pm 1$ photon *m* numbers changes this method is not effective for phase estimation when the state $|\tilde{\psi}\rangle$ inside the interferometer is an entangled or any other state in which the difference in its *m* numbers is larger than 1.

**B. Phase estimation by parity measurements of photon counting ('Second method')**

Many works [23-31] have analyzed the possibility to choose $\hat{O}$ as the parity operator $\hat{P}$ operating on one output mode. We can use photon counting for realizing measurements of the parity operator. The photon numbers distribution in the MZI output can be given as

$$|\psi\rangle_{out} = \sum_{l(1),l(2)=0}^{\infty} C_{l(1),l(2)}(out) |l(1),l(2)\rangle \quad , \tag{17}$$

where $C_{l(1),l(2)}(out)$ are the amplitudes for the photon numbers $l(1)$ and $l(2)$ in the first and second output ports of the MZI, respectively. We define the output photon numbers operators as

$$\hat{l}(1) = \left(\hat{a}^\dagger \hat{a}\right)_{out} \quad ; \quad \hat{l}(2) = \left(\hat{b}^\dagger \hat{b}\right)_{out} \quad . \tag{18}$$

Then by using $\hat{O} = \hat{P} = (-1)^{\hat{a}^\dagger \hat{a}}$ we get

$$\langle \psi_{out} | \hat{P} | \psi_{out} \rangle = \langle \psi_{out} | (-1)^{\hat{a}^\dagger \hat{a}} | \psi_{out} \rangle = \sum_{l(1),l(2)=0}^{\infty} |C_{l(1),l(2)}(out)|^2 (-1)^{l(1)} \tag{19}$$

By many photon counting measurements, giving in an histogram the number of measurements as function of $l(1)l(2)$ the probabilities $|C_{l(1),l(2)}(out)|^2$ can be obtained. The multiplication by $(-1)^{l(1)}$ of each probability can be done numerically, but one should take into account that $(-1)^{l(1)}$ depends critically on accurate measurement of $l(1)$. Photon counting methods for parity measurements are efficient therefore usually for low photon numbers.

While the phase estimation method by using Eq. (11) is quite convenient for practical calculations, it does not address the following problem:
What is the minimal phase uncertainty which we will be able to measure for a certain quantum state? This problem has been solved and analyzed in [1-4], by using quantum Fisher information, generalizing the classical Cramer-Rao relations to quantum mechanics. We review here , shortly, the answer to this problem, as we compare our calculations with certain results from these theories for pure quantum states.

In general measurements theories [1-4] one considers a curve $\hat{\rho}(X)$ on the space of density operators which are functions of the classical parameter $X$. The problem of distinguishing $\hat{\rho}(X)$ from neighboring density operators along this curve is equivalent to the problem of determining the value of the parameter $X$ [4]. The most general measurement permitted by quantum mechanics [32] is described by a set of nonnegative, Hermitian operators $\hat{E}(\xi)$ which are complete in the sense that

$$\int d\xi \hat{E}(\xi) = 1 \quad (unit\ operator) \quad . \tag{20}$$

The quantity $\xi$ labels the "results" of measurements. The probability $\hat{E}(\xi)d\xi$ represents a "positive-operator-valued-measure " (POVM). The probability distribution for the result $\xi$, given the parameter $X$, is given by

$$p(\xi | X) = Tr(\hat{E}(\xi)\hat{\rho}(X)) \quad . \tag{21}$$

By using the POVM properties $p(\xi | X)$ becomes a normalized probability distribution function.

We are interested especially in the quantum mechanical (QM) bound on the measurement of $\varphi$ in MZI . Then the density operator inside the interferometer is described by $\hat{\rho}(\varphi)$ , i.e. , $X = \varphi$ , where $\varphi$ is the classical phase parameter. By developing the general quantum measurement theory [1-4] it has been shown that the lowest possible phase uncertainty measurement which is attainable by parameter estimation is inversely proportional to the square root of the Fisher information. The quantum information $F_Q$ depends only on the state of the system and is a function of the density operator $\hat{\rho}(\varphi)$. Optimization over measurements yields the quantum bound, representing a generalization of the classical Cramer-Rao bound, given as

$$(\delta\varphi)^2 \geq \frac{1}{F_Q} \quad , \tag{22}$$

where $F_Q$ is given by

$$F_Q = Tr\left[\hat{\rho}(\varphi)\hat{A}^2\right] \quad . \tag{23}$$

Here the Hermitian operator $\hat{A}$ is called the "symmetric logarithmic derivative" (SLD) and is defined via the relation

$$\frac{\partial \hat{\rho}(\varphi)}{\partial \varphi} = \frac{1}{2}\left[\hat{A}\hat{\rho}(\varphi) + \hat{\rho}(\varphi)\hat{A}\right] \quad . \tag{24}$$

In the eigenvalue basis of $\hat{\rho}(\varphi)$, $\hat{A}$ is given by

$$(A)_{i,j} = \frac{2}{p_i + p_j}\left[\hat{\rho}'(\varphi)\right]_{i,j} \quad , \tag{25}$$

where $\hat{\rho}'(\varphi) = \frac{\partial \hat{\rho}(\varphi)}{\partial \varphi}$ and $p_i, p_j$ are the eigenvalues of $\hat{\rho}(\varphi)$ (For $p_i + p_j = 0$ we set $(A)_{i,j} = 0$).

The diagonalization of $\hat{\rho}(\varphi)$ might turn to be quite complicated. For the case of a pure quantum quantum state (no losses) the quantum Fisher information reads [1-4, 33]

$$F_Q = 4\left[\langle\psi'(\varphi)|\psi'(\varphi)\rangle - |\langle\psi'(\varphi)|\psi(\varphi)\rangle|^2\right] \quad . \tag{26}$$

For the MZI, $|\psi(\varphi)\rangle$ of Eq. (26) is the quantum state before BS2, i.e., $|\psi(\varphi)\rangle \to |\tilde{\psi}(\varphi)\rangle$, and the prime indicates a derivative with respect to $\varphi$. Since we treat in the present article pure quantum states Eqs. (22) and (26) give the minimal phase uncertainty attainable by the phase estimation theories.

The use of Eqs. (22) and (26) for obtaining the minimal phase uncertainty for a pure quantum state can be related to 'Hilbert space metric' (see e.g. [34]). The distance *dL* between two quantum states $|\psi(s)\rangle$ and $|\psi(s')\rangle$ represents a dependence on parameter $s$ and $s'$, respectively, and can be given as

$$dL^2 = 1 - |\langle \psi(s)|\psi(s')\rangle|^2 \quad . \tag{27}$$

where for $s = s'$ the distance $dL$ vanishes. After some algebra [34] one gets

$$\left(\frac{dL}{ds}\right)^2 = \left\langle \frac{d\psi(s)}{ds} \bigg| \frac{d\psi(s)}{ds} \right\rangle - \left\langle \frac{d\psi(s)}{ds} \bigg| \psi(s) \right\rangle \left\langle \psi(s) \bigg| \frac{d\psi(s)}{ds} \right\rangle \quad . \tag{28}$$

By substituting $s = \varphi$, Eq. (28) becomes equivalent to $F_Q$ (up to a factor 4) which is inversely proportional to the minimal attainable phase uncertainty squared $(\delta\varphi)^2$.

The derivation of Eq. (28) can be obtained also in another way. The *horizontal component* of the tangent vector $\frac{d\psi(s)}{ds}$ is given by $\frac{d\psi(s)}{ds} - \left\langle \psi(s)\big|\frac{d\psi(s)}{ds}\right\rangle |\psi(s)\rangle$. Here we have substracted from the derivative of the wavefunction its movement along its 'fibre' (i.e., the ray). The squared norm of the horizontal component of the wavevector is given after a straightforward algebra [34] as

$$\left\{ \left\langle \frac{d\psi(s)}{ds}\bigg| - \left\langle \frac{d\psi(s)}{ds}\bigg|\psi(s)\right\rangle\langle\psi(s)|\right\}\left\{\bigg|\frac{d\psi(s)}{ds}\right\rangle - \left\langle\psi(s)\bigg|\frac{d\psi(s)}{ds}\right\rangle|\psi(s)\rangle\right\} = 2\left(\frac{dL}{ds}\right)^2 , \tag{29}$$

which is equivalent to $F_Q$ (up to a factor 2). One should take into account that small differences in $(\delta\varphi)$ might follow from small differences in its definition.

The present paper is arranged as follows:

We use the 'First method' for calculating in Section 2 the phase uncertainty for input coherent states in the interferometer. As expected we get under *optimal conditions* the SQL phase uncertainty. It is, however, important to show under what condition the SQL is obtained even in this relatively simple case. In Section 3 we use this method for calculating the phase uncertainty for an input Fock number state in one input port, and the vacuum in the other input port, and with a 50:50 BS1. It will be interesting to note that the SQL is obtained also here in spite of the fact that we use in this Section the *nonclassical* input number state $|N\rangle$. We also show in this Section that the 'First method' gives vanishing signal for twin Fock state $|N\rangle_a|N\rangle_b$ inserted into the two input ports of the 50:50 BS1, (in agreement with the claim made in [19]). In Section 4 we calculate by the same method the optimal phase uncertainty for inserting squeezed-vacuum and coherent state into the two input ports of 50:50 BS1, respectively. It is shown that under 'optimal conditions' the phase uncertainty is below the SQL (up to a factor $e^{-r}$ where $r$ is the squeezing parameter). As the mathematical analysis for MZI is similar to that of Michelson interferometer (MI), the results are in agreement with those obtained by other methods for MI (see e.g. [35-37]).

In Section 5 we discuss the use of parity measurement by photon counting ('Second method') for phase estimation for the case where the state $|\tilde{\psi}\rangle$ inside the interferometer is a NOON state. Parity measurements of entangled states are very sensitive to losses and we would like to show that the effects of losses in such measurements can be decreased significantly by using the photon counting methods.

In Section 6 we calculate the minimal phase uncertainty obtained by using Eqs. (22) and (26), for the examples treated in Sections (2,3,5), and show that the results are in agreement with the phase uncertainty obtained by using Eq. (11). For the case of Section 4 a similar agreement is obtained after very lengthy and tedious calculations where for simplicity of presentation the detailed calculations are not given here.

Photon counting data can be used also for obtaining other photon counting moments and other photon statistics measurements. We limit the discussion, however, for the above cases which seem to be relatively easier for implementing phase estimation. Inclusion of losses in MZI or MI can be described by inserting fictitious beam splitters of transmittivity $\eta_a, \eta_b$ in the arms $a$ and $b$ of the interferometer, respectively, so that the radiation which is not transmitted, will represent losses. We treat only pure quantum states but refer to the literature [33,38,8] for generalizing the present models to include losses. We refer also to the literature on the possibility to produce entangled states conditioned on photo-detection [39]. In Section 7 we discuss the general minimal limit for phase uncertainty and summarize our results and conclusions.

## 2. PHASE UNCERTAINTY OBTAINED BY PHOTON COUNTING MEASUREMENTS FOR COHERENT STATES INPUTS IN MZI, USING EQ. (11)

The coherent states inputs after BS1 can be described as a multiplication of two coherent states:

$$|\psi\rangle = |\alpha\rangle_1 |\beta\rangle_2 = \exp\left(-\frac{|\alpha|^2}{2}\right) \sum_{n(1)=0}^{\infty} \frac{\alpha^{n(1)}}{\sqrt{n(1)!}} |n(1)\rangle \exp\left(-\frac{|\beta|^2}{2}\right) \sum_{n(2)=0}^{\infty} \frac{\beta^{n(2)}}{\sqrt{n(2)!}} |n(2)\rangle, \quad (30)$$

where $|\alpha\rangle_1$ and $|\beta\rangle_2$ are the coherent states transmitted into the first and second arms of the MZI, respectively, with corresponding annihilation operators $\hat{a}$ and $\hat{b}$ satisfying the relations

$$\hat{a}|\alpha\rangle_1 = \alpha|\alpha\rangle_1 \quad ; \quad \hat{b}|\beta\rangle_2 = \beta|\beta\rangle_2 \quad . \tag{31}$$

and where the subscripts 1 and 2 refer to the first and second arm of the MZI, respectively. Notice that we have defined here the state $|\psi\rangle$ as the state after BS1, and the use of such definition is based on the fact that BS1 produces a simple unitary transformation by which coherent states inputs in BS1 are transformed into coherent states outputs after BS1.

We assume that, by the insertion of a phase difference $\varphi$ between the two arms of the interferometer, we get the changes:

$$\begin{aligned}|\beta\rangle &= ||\beta|\exp(i\theta_2)\rangle \rightarrow |\beta'\rangle = ||\beta|\exp[i(\theta_2 + \varphi)]\rangle \quad , \\ |\psi\rangle &= |\alpha\rangle_1 |\beta\rangle_2 \rightarrow |\tilde{\psi}\rangle = |\alpha\rangle_1 |\beta'\rangle_2\end{aligned} \tag{32}$$

where we have assumed here that $|\alpha\rangle_1$ remains unchanged.

Using Eq. (16) we get

$$\begin{aligned}\langle \hat{O}_{in} \rangle &= \langle \tilde{\psi}|\hat{J}_x|\tilde{\psi}\rangle = \langle\alpha|\langle\beta'|\frac{\hat{a}^\dagger \hat{b} + \hat{b}^\dagger \hat{a}}{2}|\alpha\rangle|\beta'\rangle = \\ \frac{\alpha^*\beta' + \beta'^*\alpha}{2} &= |\alpha||\beta|\cos(\varphi + \theta_2 - \theta_1)\end{aligned} \tag{33}$$

$$\begin{aligned}\langle \hat{O}_{in}^2 \rangle &= \langle \tilde{\psi}|\hat{J}_x^2|\tilde{\psi}\rangle = \langle\alpha|\langle\beta'|\left(\frac{\hat{a}^\dagger \hat{b} + \hat{b}^\dagger \hat{a}}{2}\right)^2|\alpha\rangle|\beta'\rangle = \\ \frac{\alpha^{*2}\beta'^2 + \beta'^{*2}\alpha^2 + 2\alpha^*\alpha\beta'^*\beta' + \alpha^*\alpha + \beta'^*\beta'}{4} &= \\ \frac{|\alpha|^2|\beta|^2\left[\cos[2(\varphi + \theta_2 - \theta_1)] + 1\right]}{2} + \frac{|\alpha|^2 + |\beta|^2}{4} &= \\ |\alpha|^2|\beta|^2\cos^2[(\varphi + \theta_2 - \theta_1)] + \frac{|\alpha|^2 + |\beta|^2}{4}\end{aligned} \tag{34}$$

Substituting these relations in Eqs. (9) and (11) we get

$$\Delta \hat{O} = \sqrt{\langle \hat{O}_{in}^2 \rangle - \langle \hat{O}_{in} \rangle^2} = \sqrt{\frac{|\alpha|^2 + |\beta|^2}{4}} \quad ; \quad \left| \frac{\partial}{\partial \varphi} \langle \hat{O}_{in} \rangle \right| = |\alpha||\beta||\sin(\varphi + \theta_2 - \theta_1)| \quad ,$$

$$\delta \varphi = \frac{\Delta \hat{O}}{|\partial / \partial \varphi \langle \hat{O} \rangle|} = \frac{\sqrt{|\alpha|^2 + |\beta|^2}}{2|\alpha||\beta||\sin(\varphi + \theta_2 - \theta_1)|} \tag{35}$$

Under the assumption that $|\alpha| = |\beta|$ and that $(\varphi + \theta_2 - \theta_1) = \pi / 2$ we get for the optimal phase uncertainty $\delta \varphi = \frac{1}{\alpha \sqrt{2}} = \frac{1}{\beta \sqrt{2}}$. Since the average total number of photons $N$ used in the *symmetric* MZI is given by $N_{tot} = 2|\alpha|^2 = 2|\beta|^2$ we get $\delta \varphi = \frac{1}{\sqrt{N_{tot}}}$ which represents the SQL.

## 3. PHASE ESTIMATION BY PHOTON COUNTING FOR THE CASE IN WHICH A NUMBER STATE IS INSERTED INTO ONE INPUT PORT OF 50:50 BS1 AND THE VACUUM INSERTED IN THE OTHER INPUT PORT, BY USING EQ. (11))

Let us assume that we insert a Fock number state $\frac{(\hat{a}^\dagger)_{inp}^N}{\sqrt{N!}}|0\rangle_a$ into the input port $a$ and the vacuum into input port $b$ of MZI. By choosing a certain special phase relations for BS1 its transformation can be given as

$$|\psi\rangle_{in} = \hat{a}_{inp}^{\dagger N} |0\rangle / \sqrt{N!} |0\rangle_a \Rightarrow |\psi\rangle = \frac{(\hat{a}^\dagger + \hat{b}^\dagger)^N}{(\sqrt{2})^N \sqrt{N!}} |0\rangle_a |0\rangle_b \quad , \tag{36}$$

where all exponents of operators $\hat{a}^\dagger$ and $\hat{b}^\dagger$ operate on the vacuum states $|0\rangle_a$ and $|0\rangle_b$, respectively. We have assumed the transformation of Eq. (36), for the simplicity of calculations. The insertion of a phase difference $\varphi$ between the two arms of the interferometer transforms the state (36) into

$$|\tilde{\psi}\rangle_N = \frac{(\hat{a}^\dagger + \hat{b}^\dagger e^{i\varphi})^N}{(\sqrt{2})^N \sqrt{N!}} |0\rangle_a |0\rangle_b \quad , \tag{37}$$

where for simplicity of calculations we have assumed that the phase difference $\varphi$ is inserted in mode b. We will take into account that $|\tilde{\psi}\rangle_N$ is normalized ,i.e., $_N\langle \tilde{\psi} | \tilde{\psi} \rangle_N = 1$ for any integer N.

The expectation value for $\langle \hat{O} \rangle = {}_N\langle \tilde{\psi} | \hat{J}_x | \tilde{\psi} \rangle_N$ is given by

$$\langle \hat{O} \rangle = {}_N\langle \tilde{\psi} | \frac{\hat{a}^\dagger \hat{b} + \hat{b}^\dagger \hat{a}}{2} | \tilde{\psi} \rangle_N \quad . \tag{38}$$

We substitute Eq. (37) into Eq. (38), move operators $\hat{a}$ and $\hat{b}$ to the right, and $\hat{a}^\dagger$ and $\hat{b}^\dagger$ to the left, apply the relations $\hat{a}|0\rangle = \hat{b}|0\rangle = \langle 0|a^\dagger = \langle 0|\hat{b}^\dagger = 0$, and use the following algebraic boson commutation relations (CR):

$$\left[\hat{a}, \left(\hat{a}^\dagger + \hat{b}^\dagger e^{i\varphi}\right)^N\right]|0\rangle_a |0\rangle_b = N\left(\hat{a}^\dagger + \hat{b}^\dagger e^{i\varphi}\right)^{N-1}|0\rangle_a |0\rangle_b \ ,$$

$$\left[\hat{b}, \left(\hat{a}^\dagger + \hat{b}^\dagger e^{i\varphi}\right)^N\right]|0\rangle_a |0\rangle_b = Ne^{i\varphi}\left(\hat{a}^\dagger + \hat{b}^\dagger e^{i\varphi}\right)^{N-1}|0\rangle_a |0\rangle_b \ , \tag{39}$$

and the Hermitian conjugate of these equations.
Then by using Eqs. (37-39) we get

$$\langle \hat{O} \rangle = {}_b\langle 0|{}_a\langle 0| \frac{\left(\hat{a} + \hat{b} e^{-i\varphi}\right)^{N-1}}{\left(\sqrt{2}\right)^N \sqrt{N!}} \frac{\left(\hat{a}^\dagger + \hat{b}^\dagger e^{i\varphi}\right)^{N-1}}{\left(\sqrt{2}\right)^N \sqrt{N!}} N^2 \cos\varphi |0\rangle_a |0\rangle_b =$$

$${}_{N-1}\langle \tilde{\psi} | \tilde{\psi} \rangle_{N-1} \frac{N\cos\varphi}{2} = \frac{N\cos\varphi}{2} \quad . \tag{40}$$

The expectation value for $\langle \hat{O}^2 \rangle$ is given by

$$\langle \hat{O}^2 \rangle = {}_N\langle \tilde{\psi} | \hat{J}_x^2 | \tilde{\psi} \rangle_N = {}_N\langle \tilde{\psi} | \left(\frac{\hat{a}^\dagger \hat{b} + \hat{b}^\dagger \hat{a}}{2}\right)^2 | \tilde{\psi} \rangle_N =$$

$${}_N\langle \tilde{\psi} | \left(\hat{a}^{\dagger 2}\hat{b}^2 + \hat{b}^{\dagger 2}\hat{a}^2 + 2\hat{a}^\dagger \hat{b}^\dagger \hat{a}\hat{b} + \hat{a}^\dagger \hat{a} + \hat{b}^\dagger \hat{b}\right) | \tilde{\psi} \rangle_N \tag{41}$$

In order to apply the present method of calculation we need in addition to Eqs. (39) to use the following relations:

$$\left[\hat{a}^2, \left(\hat{a}^\dagger + \hat{b}^\dagger e^{i\varphi}\right)^N\right]|0\rangle_a |0\rangle_b = N(N-1)\left(\hat{a}^\dagger + \hat{b}^\dagger e^{i\varphi}\right)^{N-2}|0\rangle_a |0\rangle_b \ ,$$

$$\left[\hat{b}^2, \left(\hat{a}^\dagger + \hat{b}^\dagger e^{i\varphi}\right)^N\right]|0\rangle_a |0\rangle_b = N(N-1)e^{i2\varphi}\left(\hat{a}^\dagger + \hat{b}^\dagger e^{i\varphi}\right)^{N-2}|0\rangle_a |0\rangle_b \ , \tag{42}$$

and the Hermitian conjugate of these equations.
We get after somewhat lengthy but straightforward calculations :

$$\langle \hat{O}^2 \rangle =$$

$$_{N-2}\langle \tilde{\psi} | \tilde{\psi} \rangle_{N-2} \left( \frac{N(N-1)(\cos 2\varphi + 1)}{8} \right) + {}_{N-1}\langle \tilde{\psi} | \tilde{\psi} \rangle_{N-1} \frac{N}{4} = \left( \frac{N(N-1)\cos^2 \varphi + N}{4} \right). \quad (43)$$

Using Eqs. (40) and (43) we get

$$\langle \hat{O}^2 \rangle - \langle \hat{O} \rangle^2 = \frac{N(N-1)\cos^2 \varphi + N}{4} - \frac{N^2 \cos^2 \varphi}{4} = \frac{N \sin^2 \varphi}{4} \quad . \quad (44)$$

By Eq. (11) we get the phase uncertainty $\delta\varphi$ as

$$\delta\varphi = \frac{\sqrt{\langle \hat{O}^2 \rangle - \langle \hat{O} \rangle^2}}{|\partial/\partial\varphi(N\cos\varphi/2)|} = \frac{\sqrt{N}\sin(\varphi)/2}{N\sin(\varphi)/2} = \frac{1}{\sqrt{N}} \quad , \quad (45)$$

which gives the SQL, although we have used here a *nonclassical* number state as an input into 50:50 BS1 of MZI.

For inserting twin photon state $|N\rangle_a |N\rangle_b$ in the two input ports of MZI with 50:50 BS1 we note the relation

$$\left( \hat{a}^\dagger + \hat{b}^\dagger \right)^N \left( \hat{a}^\dagger - \hat{b}^\dagger \right)^N = \left( \hat{a}^{\dagger 2} - \hat{b}^{\dagger 2} \right)^N \quad . \quad (46)$$

By doing the analysis for this case and by following the use of Eq. (46) we find that the state $|\tilde{\psi}\rangle$ inside the MZI does not include 1 photon differences so that the signal $\langle \tilde{\psi} | \hat{J}_x | \tilde{\psi} \rangle$ vanishes (in agreement with [19] showing that ordinary phase measurement by currents substraction does not work).

## 4. PHASE ESTIMATION BY PHOTON COUNTING FOR THE CASE IN WHICH A COHERENT STATE IS INSERTED INTO ONE INPUT PORT AND SQUEEZED VACUUM INSERTED IN THE OTHER INPUT PORT, BY USING EQ. (11)

Let us assume that we insert a coherent state into one input port of MZI and a squeezed vacuum state into the other input port. The quantum state before BS1 is then given by [35]:

$$|\psi_0\rangle = \hat{D}_a(\alpha)|0\rangle_a \hat{S}_b(\varsigma)|0\rangle_b \quad , \quad (47)$$

where

$$\hat{D}_a(\alpha) = \exp\left( \alpha \hat{a}^\dagger - \alpha^* \hat{a} \right) \quad , \quad \hat{S}_b(\varsigma) = \exp\left\{ \frac{1}{2}\left( \varsigma^* \hat{b}^2 - \varsigma \hat{b}^{\dagger 2} \right) \right\} \quad , \quad (48)$$

and $a$ and $b$ refere to the two arms of the interferometer with corresponding creation operators. $\hat{D}_a(\alpha)$ and $\hat{S}_b(\varsigma)$ are the displacement and squeezing operators for modes $\hat{a}$ and $\hat{b}$, respectively. The complex numbers $\alpha$ and $\varsigma$ are the coherent state and the squeezed state parameters, respectively, where

$$\varsigma = r^{i\theta} \quad (r, \theta \text{ real}) \tag{49}$$

Let us assume that we have a simple unitary transformation of BS1 which transforms operators $\hat{a}$ and $\hat{b}$ as

$$\hat{a} \to \frac{\hat{a}+\hat{b}}{\sqrt{2}} \quad ; \quad \hat{b} \to \frac{\hat{a}-\hat{b}}{\sqrt{2}} \quad , \tag{50}$$

and then we get for the state $|\psi\rangle_{in}$ inserted after BS1, into the MZI

$$|\psi\rangle_{in} = \exp\left[\alpha \frac{(\hat{a}^\dagger+\hat{b}^\dagger)}{\sqrt{2}} - \alpha^* \frac{(\hat{a}+\hat{b})}{\sqrt{2}}\right] \exp\left\{\frac{1}{2}\left[\varsigma^* \frac{(\hat{a}-\hat{b})^2}{\sqrt{2}} - \varsigma \frac{(\hat{a}^\dagger-\hat{b}^\dagger)^2}{\sqrt{2}}\right]\right\}|0\rangle_a|0\rangle_b . \tag{51}$$

For simplicity of calculations let us assume that the phase difference inserted between the two arms of MZI lead to a change in the mode $b$ as

$$\hat{b} \to \hat{b}e^{-i\varphi} \quad ; \quad \hat{b}^\dagger \to \hat{b}^\dagger e^{i\varphi} \quad , \tag{52}$$

while modes $\hat{a}$ and $\hat{a}^\dagger$ remain unchanged. Then Eq. (51) is transformed into

$$|\tilde{\psi}\rangle =$$

$$\exp\left[\alpha \frac{(\hat{a}^\dagger+\hat{b}^\dagger e^{i\varphi})}{\sqrt{2}} - \alpha^* \frac{(\hat{a}+\hat{b}e^{-i\varphi})}{\sqrt{2}}\right] \exp\left\{\frac{1}{2}\left[\varsigma^* \frac{(\hat{a}-\hat{b}e^{-i\varphi})^2}{\sqrt{2}} - \varsigma \frac{(\hat{a}^\dagger-\hat{b}^\dagger e^{i\varphi})^2}{\sqrt{2}}\right]\right\}|0\rangle_a|0\rangle_b$$

$$\tag{53}$$

By series expansion of Eq. (53) we will get complicated series multiplications of photon numbers in mode $\hat{a}$ times those of mode $\hat{b}$ which can be expressed as

$$|\tilde{\psi}\rangle = \sum_{n(1),n(2)=0}^{\infty} C_{n(1),n(2)} |n(1)\rangle_a |n(2)\rangle_b \quad . \tag{54}$$

where the amplitudes $C_{n(1),n(2)}$ will turn to be quite complicated. The phase estimation method allows to use the following simpler approach.

We introduce operators $\hat{c}$ and $\hat{d}$ defined as

$$\hat{c} = \frac{\hat{a}+\hat{b}e^{-i\varphi}}{\sqrt{2}} \quad ; \quad \hat{d} = \frac{\hat{a}-\hat{b}e^{-i\varphi}}{\sqrt{2}} \quad . \tag{55}$$

Then Eq. (53) gets the simple form

$$|\tilde{\psi}\rangle = \hat{D}_c(\alpha)\hat{S}_d(\varsigma)|0\rangle_c|0\rangle_d \quad , \tag{56}$$

where $\hat{D}_c(\alpha)$ and $\hat{S}_d(\varsigma)$ are the displacement and squeezing operators in the new operators $\hat{c}$ and $\hat{d}$, respectively.

In order to calculate the expectation value of $\hat{J}_x$ and $\hat{J}_x^2$ we need to express operators $\hat{a}$ and $\hat{b}$ as functions of $\hat{c}$ and $\hat{d}$ by using Eqs, (55). We get

$$\hat{a} = \frac{\hat{c}+\hat{d}}{\sqrt{2}} \quad ; \quad \hat{b} = e^{i\varphi}\left(\frac{\hat{c}-\hat{d}}{\sqrt{2}}\right) \quad , \tag{57}$$

and then

$$\hat{J}_x = (\hat{a}^\dagger\hat{b}+\hat{b}^\dagger\hat{a})/2 = \cos\varphi\left[\hat{c}^\dagger\hat{c}-\hat{d}^\dagger\hat{d}\right] + i\sin\varphi\left[\hat{d}^\dagger\hat{c}-\hat{c}^\dagger\hat{d}\right] \quad . \tag{58}$$

We use the following relations [35]:

$$\hat{D}_c^{-1}(\alpha)\hat{c}\hat{D}_c(\alpha) = \hat{c}+\alpha \quad ; \quad \hat{D}_c^{-1}(\alpha)\hat{c}^\dagger\hat{D}_c(\alpha) = \hat{c}^\dagger+\alpha^* \quad .$$
$$\hat{S}_d^{-1}(\varsigma)\hat{d}\hat{S}_d(\varsigma) = \hat{d}\cosh(r)-\hat{d}^\dagger e^{i\theta}\sinh(r) \quad ; \tag{59}$$
$$\hat{S}_d^{-1}(\varsigma)\hat{d}^\dagger\hat{S}_d(\varsigma) = \hat{d}^\dagger\cosh(r)-\hat{d}e^{-i\theta}\sinh(r)$$

We notice that operators $\hat{c}$ and $\hat{d}$ commute and consequently $\hat{D}_c(\alpha)$ and $\hat{S}_d(\varsigma)$ also commute. We get then the following relation:

$$\hat{S}_d^{-1}(\varsigma)\hat{D}_c^{-1}(\alpha)\hat{J}_x\hat{D}_c(\alpha)\hat{S}_d(\varsigma) =$$
$$\cos\varphi\left\{(\hat{c}^\dagger+\alpha^*)(\hat{c}+\alpha)-(\hat{d}^\dagger\cosh(r)-\hat{d}e^{-i\theta}\sinh(r))(\hat{d}\cosh(r)-\hat{d}^\dagger e^{i\theta}\sinh(r))\right\} \tag{60}$$
$$+i\sin\varphi\left\{(\hat{d}^\dagger\cosh(r)-\hat{d}e^{i\theta}\sinh(r))(\hat{c}+\alpha)-(\hat{c}^\dagger+\alpha^*)(\hat{d}\cosh(r)-\hat{d}^\dagger e^{-i\theta}\sinh(r))\right\}$$

For the expectation values of $\hat{J}_x$ we get

$$\langle 0|_d\langle 0|_c \hat{S}_d^{-1}(\varsigma)\hat{D}_c^{-1}(\alpha)\hat{J}_x\hat{D}_c(\alpha)\hat{S}_d(\varsigma)|0\rangle_c|0\rangle_d = \cos\varphi\left[|\alpha|^2-\sinh^2 r\right] \quad , \tag{61}$$

where the displacement operator has transformed only mode $\hat{c}$ and the squeezing operator has transformed only mode $\hat{d}$. In deriving Eq. (61) all the terms which are proportional to $i\sin\varphi$ vanish and only two terms remain which are proportional to $\cos\varphi$. (Eq. (61) is equivalent to Eq. (2.33a) of [35]).

In the present analysis the phase estimation operator is $\hat{J}_x$ and according to Eq. (11) we need to get a maximal value for $\left|\frac{\partial}{\partial \varphi}\langle \hat{J}_x \rangle\right|$. By assuming $|\alpha|^2 \gg \sinh^2 r$ this maximal value is obtained approximately by choosing $\cos\varphi = 0$ and then it is given by

$$\left|\frac{\partial}{\partial \varphi}\langle \hat{J}_x \rangle\right|_{\sin\varphi=1} = |\alpha|^2 - \sinh^2(r) \approx |\alpha|^2 \quad . \tag{62}$$

By assuming $\cos\varphi = 0$, and neglecting small terms which are not proportional to $|\alpha|^2$, we get after straightforward calculations

$$\langle 0|_d \langle 0|_c \left\{\hat{S}_c^{-1}(\varsigma)\hat{D}_d^{-1}(\alpha)\hat{J}_x^2 \hat{D}_d(\alpha)\hat{S}_c(\varsigma)\right\}|0\rangle_c |0\rangle_d \approx$$
$$|\alpha|^2 \left[\cosh^2(r) + \sinh^2(r)\right] + \alpha^2 \sinh(r)\cosh(r)e^{i\theta} + \alpha^{*2}\sinh(r)\cosh(r)e^{-i\theta} \quad . \tag{63}$$

Using the definition $\alpha = |\alpha|e^{if}$ and the optimal condition $\theta + 2f = \pi$ we get

$$\langle \hat{J}_x^2 \rangle \approx |\alpha|^2 \left[\cosh(r) - \sinh(r)\right]^2 = |\alpha|^2 e^{-2r} \quad . \tag{64}$$

Under the condition $\cos\varphi = 0$ we get $\langle \hat{J}_x \rangle_{\cos\varphi=0} = 0$ and then

$$\Delta\hat{O} = \sqrt{\langle \hat{O}^2 \rangle - \langle \hat{O} \rangle^2} = \left[\langle \hat{J}_x^2 \rangle - \langle \hat{J}_x \rangle^2\right]_{\cos\varphi=0,\theta+2f=\pi} \approx |\alpha|^2 e^{-2r} \quad ,$$
$$\left|\frac{\partial}{\partial \varphi}\langle \hat{J}_x \rangle\right|_{\sin\varphi=1} \approx |\alpha|^2 \tag{65}$$

Then by using Eqs. (11) and (65) we get:

$$\delta\varphi \approx \frac{|\alpha|e^{-r}}{|\alpha|^2} = \frac{e^{-r}}{|\alpha|} = \frac{e^{-r}}{\sqrt{N}} \quad . \tag{66}$$

One should notice that within our approximations the average total number of photons is approximately equal to that of the coherent state as the average number of squeezed-vacuum photons is very small ($\sinh^2 r \ll |\alpha|^2$). The squeezed-vacuum state has, however, strong fluctuations which affect the phase uncertainty. We find according to Eq. (66) that by using squeezed vacuum states the minimal value of $\delta\varphi$ is smaller by a factor $e^{-r}$ relative to the SQL. Although this result is well known one should notice that this result is obtained here only under the above approximations, and the use of phase estimation method for deriving this result should be of interest.

Similar analysis to that made for MZI can be made to MI, for gravitational waves detection, but one should take into account that in the use of very strong coherent states for detecting extremely small phase differences, other effects become important. For example, for treating radiation pressure effects see e.g. [40,41]. Also we have assumed the use of 50:50 BS1 but one can use a very special unitary transformation for Michelson interferometer in which the strong coherent state will be transmitted approximately into one output port and weak squeezed-coherent state will be transmitted into the other output 'dark' port, see [36,37]). Then, we can get the phase difference between the 'dark' port output state, and that of the approximate strong coherent state in the other output port,

## 5. PHASE ESTIMATION BY PARITY MEASUREMENT OF PHOTON COUNTING FOR A NOON STATE INSETED INTO MZI, BY USING EQ. (11)

We treat here a NOON state [42-45] where its form *inside* the interferometer after BS1 is given as

$$|\psi\rangle_{in} = (|N\rangle_a |0\rangle_b + |0\rangle_a |N\rangle_b)/\sqrt{2} \qquad (67)$$

In order to explain the nature of this state we can consider the simple special case in which one photon is inserted in each input port of 50:50 BS1 and the state exiting this BS is given as

$$|\psi\rangle_{in} = (|2\rangle_a |0\rangle_b + |0\rangle_a |2\rangle_b)/\sqrt{2} \qquad (68)$$

The output from BS1 can be expressed according to Eq. (6) as

$$|\psi\rangle_{in} = (|j=1, m=1\rangle + |j=1, m=-1\rangle)/\sqrt{2} \qquad . \qquad (69)$$

Then the state before BS2 is given according to Eq. (7) by

$$|\tilde{\psi}\rangle = (|1,1\rangle e^{-i\varphi} + |1,-1\rangle e^{i\varphi})/\sqrt{2} \qquad . \qquad (70)$$

We use a short notation in which the first number in the ket represents $j$ and the second number represents $m$. One should take into account that only the *phase difference* introduced between the two arms of the interferometer will enter in the analysis.

We are interested in generalization of Eq. (70) to NOON states with larger $j$ values, i.e., of the form

$$|\psi\rangle_{in} = (|j, j\rangle + |j, -j\rangle)/\sqrt{2} \qquad . \qquad (71)$$

There are different methods for producing NOON states (see e.g. [39-41]) but in the meantime only NOON states with small $j$ values are available. It is hoped that with new experimental developments in quantum optics it will be possible to implement such states with high $j$ values.

We assume in this Section that the state (71) is given as the *input* into MZI where BS1 is *exchanged* by a certain nonlinear physical system producing the NOON state, (or creating such state conditioned on photo detection [39]). Then, the state before BS2 is given according to Eq. (7) by

$$|\tilde{\psi}\rangle = \left(|j,j\rangle e^{-ij\varphi} + |j,-j\rangle e^{ij\varphi}\right)/\sqrt{2} \tag{72}$$

The expectation value of the parity operator, denoted as $\hat{P}$, is made over the output state exiting BS2 and is given by

$$\langle \psi|_{out} \hat{P} |\psi\rangle_{out} = \langle \psi| \hat{U}^\dagger_{BS2} (-1)^{j-\hat{J}_z} \hat{U}_{BS2} |\psi\rangle \quad , \tag{73}$$

where we substituted here $\hat{P} = (-1)^{\hat{a}^\dagger \hat{a}} = (-1)^{j-\hat{J}_z}$. We choose

$$\hat{U}_{BS2} = \exp\left[(i\pi/2)\hat{J}_x\right] \quad , \tag{74}$$

then we get [26]:

$$\hat{Q} = \hat{U}^\dagger_{BS2} (-1)^{j-\hat{J}_z} \hat{U}_{BS2} = \exp\left[(-i\pi/2)\hat{J}_x\right] (-1)^{j-\hat{J}_z} \exp\left[(i\pi/2)\hat{J}_x\right] =$$
$$(-1)^j \exp(i\pi \hat{J}_y) = \sum_{\mu,\nu=-j}^{j} (-1)^j d^j_{\nu,\mu}(-\pi) |j,\nu\rangle\langle j,\mu| \tag{75}$$

where $d^j_{\nu,\mu}$ is the rotation matrix. We use the rotation matrix equality

$$d^j_{\nu,\mu}(-\pi) = (-1)^{2\nu} \delta(\nu,-\mu) \quad , \tag{76}$$

then it is shown [26] that that the operator $\hat{Q}$ is a projection operator satisfying the relation $\hat{Q}^2 = 1$, i.e., representing the unit operator.

We get:

$$\langle \psi|_{out} \hat{O}^2 |\psi\rangle_{out} \Rightarrow \langle \psi|_{out} \hat{P}^2 |\psi\rangle_{out} = \langle \tilde{\psi}|\hat{Q}^2|\tilde{\psi}\rangle = 1 \quad . \tag{77}$$

Inserting Eq. (75) and (76) into Eq. (73) we get

$$\langle \psi|_{out} \hat{O} |\psi\rangle_{out} \Rightarrow \langle \psi|_{out} \hat{P} |\psi\rangle_{out} = \langle \tilde{\psi}|\hat{Q}|\tilde{\psi}\rangle =$$
$$\frac{\{e^{ij\varphi}\langle j,j| + e^{-ij\varphi}\langle j,-j|\}}{\sqrt{2}} \{|j,j\rangle\langle j,-j| + |j,-j\rangle\langle j,j|\} \frac{\{e^{-ij\varphi}\langle j,j| + e^{ij\varphi}|j,-j\rangle\}}{\sqrt{2}} \tag{78}$$
$$= \cos(2j\varphi)$$

In Eq. (78) the operator $\hat{Q}$ has been transformed into a simple projection operator (up to a factor $1/\sqrt{2}$) as the effects of the terms $(-1)^j$ and $(-1)^{2\nu}$ is only to introduce a possible phase term factor which can affect only the *total quantum state phase* which is not relevant to phase difference measurements in MZI (see also [8,10, 40 ] ).

In order to implement the use of Eq. (11) to parity measurements we use the following calculations:

$$\Delta \hat{O} = \sqrt{\langle \hat{O}^2 \rangle - \langle \hat{O} \rangle^2} \Rightarrow \sqrt{\langle \psi |_{out} \hat{P}^2 | \psi \rangle_{out} - \left( \langle \psi |_{out} \hat{P} | \psi \rangle_{out} \right)^2} ,$$
$$= \sqrt{1 - \cos^2(2j\phi)} = \sin(2j\varphi) \qquad (79)$$

$$\left| \frac{\partial \langle \hat{O} \rangle}{\partial \varphi} \right| = \left| \frac{\partial \langle \psi |_{out} \hat{P} | \psi \rangle_{out}}{\partial \varphi} \right| = \left| \frac{\partial}{\partial \varphi} \left[ \cos(2j\varphi) \right] \right| = |2j \sin(2j\varphi)| \qquad . \qquad (80)$$

Then by assuming the measurement operator to be the parity operator for the NOON states *inside* the interferometer, and inserting Eqs. (79) and (80) into Eq. (11), we get:

$$\delta\varphi = \frac{1}{2j} = \frac{1}{N} \quad ; \quad 2j = n_1 + n_2 = N \qquad . \qquad (81)$$

Since for the NOON state the total number of photons $N$ is given by $2j$ (see Eq. (6)), we find that the above use of NOON state in MZI implements precision which is of order $1/N$.

In the above use of NOON state for precision phase difference measurements in MZI we should notice the following facts:

a) An ordinary BS1 will not produce a NOON state. By inserting a NOON state into BS1 we will get a quite complicated entangled state. We have simplified the analysis by assuming that BS1 is *exchanged* with a nonlinear physical system producing the NOON state.

b) In the discussion after Eq. (19) we have shown that parity measurements depend critically on accurate measurement of the factor $(-1)^{l(1)}$, which is strongly "spoiled" by losses effects. Such unavoidable losses effects can be decreased by the method of photon counting eliminating in the analysis all "spoiled" measurements for which $l(1) + l(2)$ (both are measured by photon counting) are not equal to $N = 2j$, for a particular NOON state.

# 6. COMPARISONS BETWEEN THE PHASE ESTIMATIONS OBTAINED BY USING EQ. (11) FOR THE EXAMPLES OF SECTIONS 2,3,5, AND THE MINIMAL PHASE ESTIMATIONS OBTAINED BY USING EQS. (22) and (26

We have shown that phase estimation in MZI can be made by choosing a photon counting observable $\hat{O}$, which is phase dependent, and estimate the phase uncertainty by using Eq. (11) . While the photon counting measurements are made on the output of the MZI by using a special unitary transformation for BS2 the calculations can be made on the quantum state $|\tilde{\psi}\rangle$ of the EM field in MZI before BS2. Such calculations for pure quantum states have been made for some examples in Sections (2-5). It will be of interest to compare the results obtained in some examples with corresponding calculations for the minimal phase uncertainty calculated by the quantum Fisher information [1-4,36], and such comparisons are made for Sections 2,3,5 as follows (The calculations for the example of Section 4 give similar results, but these calculations are very lengthy and tedious and for the simplicity of presentation are not given here):

## A. Minimal phase-uncertainty for coherent states inputs in MZI obtained by using Eqs. (22) and (26), compared to that calculated by Eq. (11)

For the coherent states inputs in the MZI , the minimal phase uncertainty measurement can be obtained by using Eqs. (22) and (26), where in these equations we substitute according to Eqs. (30-32):

$$|\tilde{\psi}\rangle = |\alpha\rangle_1 |\beta'\rangle_2 = |\alpha\rangle_1 \sum_{n(2)=0}^{\infty} C_{n(2)} |n(2)\rangle \;\; ; \;\; C_{n(2)} = \exp\left(-\frac{|\beta|^2}{2}\right) \frac{[|\beta|\exp[i(\theta_2 + \varphi)]]^{n(2)}}{\sqrt{n(2)!}}, \quad (82)$$

and the derivative of $|\tilde{\psi}\rangle$ operates only $|\beta'\rangle_2$ as:

$$|\tilde{\psi}'\rangle \equiv \frac{\partial |\tilde{\psi}\rangle}{\partial \varphi} = |\alpha\rangle_1 \sum_{m=0}^{\infty} \frac{\partial C_{n(2)}}{\partial \varphi} |n(2)\rangle = |\alpha\rangle_1 \sum_{n(2)=0}^{\infty} in(2) C_{n(2)} |n(2)\rangle \quad . \quad (83)$$

Then we get

$$\langle \tilde{\psi}'|\tilde{\psi}'\rangle = \sum_{n(2)=0}^{\infty} |C_{n(2)}|^2 n(2)^2 = \langle n(2)^2\rangle \;\; ; \;\; |\langle \tilde{\psi}'|\tilde{\psi}\rangle|^2 = \left[\sum_{n(2)=0}^{\infty} |C_{n(2)}|^2 n(2)\right]^2 = \langle n(2)\rangle^2 \;. \;\;(84)$$

Substituting Eqs. (84) into (22) and (26) we get

$$F_Q = 4\left[\langle n(2)^2\rangle - \langle n(2)\rangle^2\right] = 4\langle n(2)\rangle = 4|\beta|^2 \;\;,\;\; \delta\varphi = \frac{1}{2|\beta|} \quad . \quad (85)$$

In Section (2) we obtained by the use of Eq. (11) under optimal conditions $\delta\varphi = \frac{1}{\sqrt{N_{tot}}} = \frac{1}{\sqrt{2}|\beta|}$ which is larger by a factor $\sqrt{2}$ relative to that obtained in (85). This small difference might result from MZI properties or definitions.

**B. Minimal phase-uncertainty obtained by inserting a number state in one input port and the vacuum in the other input port of MZI using Eqs. (22) and (26), compared to that calculated by Eq. (11)**

Using Eq. (37), derivatives according to $\varphi$, and the CR of Eq. (39), we get after straight forward calculations:

$$\langle\tilde{\psi}'|_N \equiv \frac{\partial\langle\tilde{\psi}|_N}{\partial\varphi} = -i\sqrt{\frac{N}{2}}e^{-i\varphi}\langle\tilde{\psi}|_{N-1}\hat{b} \quad ; \quad |\tilde{\psi}'\rangle_N \equiv \frac{\partial|\tilde{\psi}\rangle_N}{\partial\varphi} = ie^{i\varphi}\sqrt{\frac{N}{2}}\hat{b}^\dagger|\tilde{\psi}\rangle_{N-1} \quad , \quad (86)$$

$$\hat{b}|\tilde{\psi}\rangle_N = \sqrt{\frac{N}{2}}e^{i\varphi}|\tilde{\psi}\rangle_{N-1} \quad ; \quad \langle\tilde{\psi}|_N \hat{b}^\dagger = \sqrt{\frac{N}{2}}e^{-i\varphi}\langle\tilde{\psi}|_{N-1} \quad . \quad (87)$$

Then we get

$$\left|\langle\tilde{\psi}'|_N|\tilde{\psi}\rangle_N\right|^2 = \left|\frac{\partial\langle\tilde{\psi}|_N}{\partial\varphi}|\tilde{\psi}\rangle_N\right|^2 = \frac{N^2}{4} \quad , \quad (88)$$

$$\langle\tilde{\psi}'|_N|\tilde{\psi}'\rangle_N \equiv \frac{\partial\langle\tilde{\psi}|_N}{\partial\varphi}\frac{\partial|\tilde{\psi}\rangle_N}{\partial\varphi} = \frac{N}{2}\langle\tilde{\psi}|_{N-1}\hat{b}\hat{b}^\dagger|\tilde{\psi}\rangle_{N-1} =$$
$$\frac{N}{2}\langle\tilde{\psi}|_{N-1}|\tilde{\psi}\rangle_{N-1} + \frac{N}{2}\langle\tilde{\psi}|_{N-1}\hat{b}^\dagger\hat{b}|\tilde{\psi}\rangle_{N-1} = \frac{N}{2} + \frac{N(N-1)}{4} = \frac{N^2+N}{4} \quad . \quad (89)$$

Using the above equations we get for the quantum Fisher information of Eq. (26):

$$F_Q = 4\left[\langle\psi'(\varphi)|\psi'(\varphi)\rangle - |\langle\psi'(\varphi)|\psi(\varphi)\rangle|^2\right] = 4\left[\frac{N}{4} + \frac{N^2}{4} - \frac{N^2}{4}\right] = N \quad (90)$$

For the phase uncertainty of Eq. (22) we get

$$(\delta\varphi) \geq \frac{1}{\sqrt{N}} \quad (91)$$

Here the optimal limit is equal to that derived in Section 3.

**C. Minimal phase-uncertainty by NOON states inputs into MZI obtained by using Eqs. (22) and (26), compared to that calculated by Eq. (11)**

We calculate here the quantum Fisher information $F_Q$ for the case in which we have the entangled NOON state $|\tilde{\psi}\rangle_{NOON}$ before BS2. By substituting the state $|\tilde{\psi}\rangle$ of Eq. (72) into Eq. (26) we get

$$\langle \tilde{\psi}'(\phi)|\tilde{\psi}'(\phi)\rangle =$$

$$\frac{\{ije^{ij\varphi}\langle j,j|+(-ij)e^{-ij\varphi}\langle j,-j|\}}{\sqrt{2}} \frac{\{-ije^{-ij\varphi}|j,j\rangle + ije^{ij\varphi}|j,-j\rangle\}}{\sqrt{2}} = j^2 \quad , \tag{92}$$

$$\langle \tilde{\psi}'(\phi)|\tilde{\psi}(\phi)\rangle =$$

$$\frac{\{ije^{ij\varphi}\langle j,j|+(-ij)e^{-ij\varphi}\langle j,-j|\}}{\sqrt{2}} \frac{\{e^{-ij\varphi}|j,j\rangle + e^{ij\varphi}|j,-j\rangle\}}{\sqrt{2}} = 0 \tag{93}$$

Substituting these results into Eq. (26) we get:

$$F_Q = 4j^2 \quad , \tag{94}$$

And then according to Eq. (22) we get

$$(\delta\varphi)^2 \geq \frac{1}{4j^2} \quad , \tag{95}$$

and the minimal phase uncertainty $(\delta\varphi)_{\min}$ which can be obtainable in experiments is given by

$$(\delta\varphi)_{\min} = \frac{1}{2j} = \frac{1}{N} \quad , \tag{96}$$

where $2j = N_1 + N_2 = N$ for NOON state. Here the optimal limit is equal to that derived for NOON state in Section 5.

## 7. SUMMARY, DISCUSSION AND CONCLUSION

In the present work we have analyzed the use of phase estimation for MZI. Photon counting is chosen in the present analysis as the observable $\hat{O}$ to be measured. While the experimental photon counting measurements are performd on the output states of the MZI denoted as $|\psi\rangle_{out}$, the phase estimation calculations are made on the states $|\tilde{\psi}\rangle$ (see Eq. (7)) which are the states before BS2. $|\psi\rangle_{out}$ is obtained from $|\tilde{\psi}\rangle$ by using a special unitary transformation for BS2.

Two photon counting methods have been developed giving phase uncertainty by using Eq. (11): a) Phase estimation is made by measuring the output photon number difference. This method has been applied to the calculations for phase uncertainty for three cases:1) For coherent states input into the MZI. 2) For the case in which a Fock number state is inserted in one input port and the vacuum in the other input port . 3) For the case in which a coherent state is inserted in one input port and squeezed-vacuum in the other input port. While for cases (1) and (2) a phase uncertainty has been calculated which is of the order of SQL, for case (3), by choosing certain optimal parameters, the calculated phase uncertainty becomes below the SQL and under optimal conditions the phase uncertainty can be reduced by a factor $e^{-r}$ relative to the SQL . b) Phase estimation by photon counting has been analyzed for a NOON state where BS1 is *exchanged* by a nonlinear system which inserts the NOON state, as an input state $|\psi\rangle_{in}$, into the interferometer, and by parity measurements. Phase uncertainty obtained in this case is of order 1/N and it has been shown that photon counting method can decrease photon losses effects.

While the above results are in agreement with expected results for such cases, the methods developed in the present analyis by using photon counting measurements in MZI are different from the usual conventional ones.

By using certain properties of 'Hilbert space metric' the minimal phase uncertainty, which can be obtained for any quantum state, has been derived in previous works [1-4]. For *a pure quantum state* the quantum Fisher information $F_Q$ is given by Eq. (26), and the minimal phase uncertainty is inversely proportional to $F_Q$, as given by Eq. (22). By using these equations, it has been shown in the present analysis, that the minimal phase uncertainty for the cases treated in Sections (2,3,5) is equal to the phase uncertainty measured by photon counting which is calculated by the use of Eq. (11).

By substituting the general state $|\tilde{\psi}\rangle$ of MZI which is before BS2 and which is given by Eq. (7), into Eq. (26), we get the following expression for $F_Q$:

$$F_Q = 4\left[\left\{\sum_{j=0}^{\infty}\sum_{m,m'=-j}^{j} C^*_{j,m}C_{j,m'}(m'-m)^2\right\} - \left\{\sum_{j=0}^{\infty}\sum_{m,m'=-j}^{j} C^*_{j,m}C_{j,m'}(m'-m)\right\}^2\right] . \quad (97)$$

This expression can be related to the operator $\hat{J}_z$ as:

$$F_Q = 4\left[\langle\tilde{\psi}|\hat{J}_z^2|\tilde{\psi}\rangle\right] - \langle\tilde{\psi}|\hat{J}_z|\tilde{\psi}\rangle^2] \quad (98)$$

According to Eq. (22) then we get

$$\delta\varphi = \frac{1}{\sqrt{F_Q}} = \frac{1}{2\sqrt{[\langle\tilde{\psi}|\hat{J}_z^2|\tilde{\psi}\rangle] - \langle\tilde{\psi}|\hat{J}_z|\tilde{\psi}\rangle^2]}} = \frac{1}{2\sqrt{[\langle(m'-m)^2\rangle - \langle m'-m\rangle^2]}} \quad , \quad (99)$$

which has the form of uncertainty relation

$$\delta\varphi 2\Delta m = 1 \quad , \quad (100)$$

but the uncertainty $\delta\varphi$ is for a classical phase difference parameter $\varphi$. (For the relation between classical and non-classical precision quantum metrology see e.g. [46]).

For the case in which the total number of photons $n_1 + n_2 = 2j$ is constant we get the limit for minimal phase uncertainty $\delta\varphi_{\min}$ by assuming $\langle\tilde{\psi}|\hat{J}_z|\tilde{\psi}\rangle = 0$ and $\langle\tilde{\psi}|\hat{J}_z^2|\tilde{\psi}\rangle = j^2$. Then:

$$\delta\varphi_{\min} = \frac{1}{2j} = \frac{1}{n_1 + n_2} = \frac{1}{N_{tot}} \qquad . \tag{101}$$

Finally I would like to point out that in another approach to phase measurements in MZI, which is different from the phase estimation approach, there is a quasi-conjugate uncertainty relation between the 'relative phase operator' and the operator representing the difference in the number of photons and *the complete quantum phase distribution can be obtained* (See for this approach in [47], and in the long list of References included). In the present work we have followed, however, the 'phase estimation methods' in which $\varphi$ is a classical parameter and it has been shown that by using photon counting measurements these methods become very efficient for deriving the two-mode phase uncertainty in MZI.